# Comments on

# ""Hot" electrons in metallic nanostructures - non-thermal carriers or heating?", Cond-Matt arXiv:1810.00565, versions 2 and 3, Light: Science & Applications, 8, 89 (2019).

# and

# "Assistance of metal nanoparticles to photo-catalysis – nothing more than a classical heat source", Faraday Discussions 214, 215–233 (2018).

Alexander O. Govorov[1] and Lucas V. Besteiro[1,2,3]

[1] *Department of Physics and Astronomy, Ohio University, Athens OH 45701, USA*

[2] *Institute of Fundamental and Frontier Sciences, University of Electronic Science and Technology of China, Chengdu 610054, China*

[3] *Centre Énergie Matériaux et Télécommunications, Institut National de la Recherche Scientifique, Varennes, Québec J3X 1S2, Canada*

[*] E-mails: govorov@ohio.edu; lucas.besteiro@emt.inrs.ca

In this version of the preprint, we made a few changes and additions. We highlight the changes with color.

**The authors of preprint [1] and paper in Light: Science & Applications (2019), Y. Dubi and Y. Sivan, made several wrong and inconsistent comments on our papers [3-5]. In addition, paper [2] by the same authors also contains several wrong statements in relationship with our work [3]. Moreover, the authors of Ref. [1] address in their comments features that were in fact not present in Refs. [3-5]. In what follows we present correction to a number of specific points in which they either misrepresented or erroneously interpreted our published work.**

When comparing our paper [3] and papers [1] and [2], we need to first clarify that papers [1] and [2] use <u>fundamentally</u> different approximations than our paper [3] does. Ref. [1] does not use quantized states and it can be applied only to relatively large NCs. Whereas our paper [3] (as well as our previous papers) uses a fully quantum set of electronic wavefunctions in a nanosphere, a nanocube and a nanosphere dimer. Therefore, the results for small NCs should be very different in papers [1] and [2]. Our paper [2] shows perfectly the quantum-to-classical transition to the limit of large NC sizes. Indeed, for large NCs, we obtain the same result as if described by the classical Drude dielectric function (the Drude limit).

In our opinion, the approach used in Ref. [1] has no conceptual novelty. However, the goal of these notes is not an evaluation of the results presented in Ref. [1], but to address the comments on our work, made in Refs. [1] and [2], and show their incorrectness.

Furthermore, we will respond to another feature of the preprint [1], which is the following: the authors criticize points that do not exist in our paper at all. For example, they commented on the electron temperature ($T_e$), when we did not discuss it at all in Ref. [3], since our approach is more general and our methods go beyond the $T_e$ approximation. At the same time, we have to note that the $T_e$ approach is not only very productive, but also highly appreciated in the literature. Regarding the study presented in Ref. [1], the use of the concept of $T_e$ in that study is not novel, in our opinion.

Now, we will comment on the specific statements from the preprint [1] (versions v2 and v3) and from the paper [2] (Faraday Discussions) that directly pertain to our published work.

# 1. Cond-matt arXiv preprint (Ref. [1]), v3, page 5, paragraph 3

Paper [1] reads: "What remains to be done is to determine $T_{ph}$ - it controls the rate of energy transfer from the electron subsystem to the phonon subsystem, and then to the environment. Recent studies of the steady-state non-equilibrium in metals (e.g., [26-28] *{[3-5] in this Commentary}*) relied on a fixed value for $T_{ph}$ (choosing it to be either identical to the electron temperature, or to the environment temperature) and/or treated the rate of *e-ph* energy transfer using the relaxation time approximation with a *e-ph* collision rate which is independent of the field and particle shape. While phenomenologically correct, these approaches ignore the dependence of the energy transfer to the environment on the nanoparticle shape, the thermal properties of the host material, the electric field strength and the temperature difference. Therefore, not only these phenomenological approaches fail to ensure energy conservation, but they also fail to provide a correct quantitative prediction of the electron distribution near the Fermi energy (which is strongly dependent on $T_{ph}$) and provides incorrect predictions regarding the role of nanoparticle shape and host properties on the steady-state electron distribution and the temperatures."

This paragraph (found in Ref. [1], v3, page 5, paragraph 3) contains several misrepresentations of our work, which we shall discuss separately below. Furthermore, we would like to express our puzzlement regarding the statements "While phenomenologically correct" and "not only these phenomenological approaches fail to ensure energy conservation", as they seem to be at odds: in our opinion, a description that does not conserve energy cannot be phenomenologically correct.

**a) Conservation of energy in Ref. [3].** Contrary to the statement above, our approach in Ref. [3], also used in our previous papers [4,5], does conserve energy. Our approach relies on the commonly-used kinetic rate equation, which conserves the total energy in all instances and regimes. Let us briefly review the kinetic equation from our paper [3] and derive from it the fundamental energy-balance equation using simple, fundamental steps.

Eq. 25 from Ref. [3] reads

$$\frac{\partial \rho_{nn}}{\partial t} = G_n - R_n$$

$\rho_{nn} \equiv$ the electron population

$G_n \equiv$ the optical excitation rate (1c)

$R_n = R_{e-e} + R_{e-phonon} \equiv$ the relaxation rate due to the e-e and e-phonon collisions

$$R_{e-phonon} = \frac{\rho_{nn} - f_F^0}{\tau_{\varepsilon, phonons}}$$

Then, we multiply both sides by the single electron energy and sum up over all states, keeping in mind that the *e-e* scattering conserves the electronic energy (i.e., $\sum_n \varepsilon_n R_{e-e} = 0$):

$$\sum_n \varepsilon_n \frac{\partial \rho_{nn}}{\partial t} = \sum_n \varepsilon_n G_n - \sum_n \varepsilon_n R_n$$

$$\frac{\partial E_e}{\partial t} = Q_{optical\ absorption} - Q_{e-phonon}$$

$E_e = \sum_n \varepsilon_n \rho_{nn} \equiv$ the total electronic energy (2c)

$Q_{optical\ absorption} \equiv$ the rate of absorption of light energy

$Q_{e-phonon} \equiv$ the rate of trasfer of energy from the electrons to phonons

$$Q_{e-phonon} = \frac{E_e - E_{e,F}(T_{lattice})}{\tau_{\varepsilon, phonons}}$$

The above equations account for the electronic system absorbing the photon energy and then transferring it to the lattice. Therefore, the energy is not only conserved, but fundamental in the construction of the formalism. Of course, the lattice's energy of a NC can then diffuse to the environment via phonon transfer and diffusion, which is a common sense assumption of the model.

**b) The role of lattice temperature.** Paper [1] reads: "Recent studies of the steady-state non-equilibrium in metals (e.g., [26-28] *{[3-5] in this Commentary}*) relied on a fixed value for $T_{ph}$ (choosing it to be either identical to the electron temperature, or to the environment temperature) and/or treated the rate of *e-ph* energy transfer using the relaxation time approximation with a *e-ph* collision rate which is independent of the field and particle shape."

The above sentence has several important and fundamental mistakes:

i) The lattice temperature ($T_{lattice}$) in Ref. [3] was taken as a parameter, which need not to be assumed to be equal to the matrix (ambient) temperature, $T_0$. Actually, it cannot be taken as exactly equal to $T_0$, fundamentally. However, and because we worked in the linear regime, we can safely approximate $T_{lattice}$ to $T_0$.

Therefore, the rates of hot-electron generation and optical absorption were calculated at room temperature, consistently with the small intensities used in our paper [3]. This approach is a common one and fully justified.

ii) We did not introduce the electronic temperature ($T_e$) in any of the cited papers (Refs. [3-5]), so the authors of Ref. [1] seem to be responding to a point of their own creation.

iii) The comment "the rate of *e-ph* energy transfer using the relaxation time approximation with a *e-ph* collision rate which is independent of the field and particle shape", extracted from the paragraph above, looks strange and unclear. Although the model does indeed use a fixed *e-ph* collision rate, $\tau_{e-ph}$, this is a parameter that, within our formalism, it can depend on the shape and size of a NC, but it should not depend on the light intensity. However, the relevant value to consider, which captures these dependences, is the relaxation rate, $R_{e-phonon}$. From Eq. (2c) we can see that in the steady-state the rate of energy transfer to phonons is equal to the absorption of light by the particle:

$$\frac{\partial E_e}{\partial t} = 0$$
$$Q_{optical\ absorption} = Q_{e-phonon}$$

The energy transfer rate from electrons to phonons is not fixed at all, as it must not be fixed for fundamental reasons.

iv) Extending the previous comment, it is worth noting that we did not argue, at all, for the rate of energy transfer to phonons being independent of field and particle shape. On the contrary, $Q_{e-phonon}$ depends on these as a fundamental property, since the absorption $Q_{optical\ absorption}$ strongly depends on the field strength and particle's size and shape. This mistake appears to reveal a fundamental misunderstanding of our model by the authors of Ref. [1].

**c) On quantitative predictions and particle's material and shape dependence.** Their comment in [1] continues: "Therefore, not only these phenomenological approaches fail to ensure energy conservation, but they also fail to provide a correct quantitative prediction of the electron distribution near the Fermi energy (which is strongly dependent on $T_{ph}$) and provides incorrect predictions regarding the role of nanoparticle shape and host properties on the steady-state electron distribution and the temperatures."

The energy is conserved in our formalism, as shown above. The phonon energy from the NC flows to the matrix, and this is the commonly-assumed approach used in our paper.

Clearly, our paper [3] presents quantitative predictions for the low-energy distributions made within the fully quantum model, and their claim that these results are not correct are, in our opinion, unsubstantiated. This is so because they base such claim on premises that, as we have shown above, do not resist scrutiny. Furthermore, a disagreement between the results presented in Ref. [1] and Ref. [3] does not warrant such conclusion either, given the important differences of the models used: The paper [1] does not consider the quantization of the electronic states and can be therefore applied only to large NCs. Moreover, Ref. [1] uses a bulk mechanism for the excitation of electrons that ignores the size quantization effects. In contrast, our paper [3] provides a set of valid predictions for the absorptions, rates and electron distributions, arising from a physically-justified model that, importantly, relies on a quantum description of the electronic system. Furthermore, the lattice temperature of a NC was taken as a parameter, but it can be estimated or calculated using well-known and well-established mathematical approaches [6,7,8]. Therefore, our paper [3] has no "incorrect predictions regarding the role of nanoparticle shape and host properties" and the paper [1] includes critical comments that do not make sense. To conclude regarding our study, our paper [3] presented an original theoretical study of the hot-electron generation effect, obtained within a consistent quantum formalism, and which we believe will be useful for the field.

# 2. Cond-mat preprint (Ref. [1]), v3, page 3, paragraph 1.

Referring to Refs. [4,5], a comment in Ref. [1], on page 3, reads: "… under CW illumination [26,27] accounted for the electron distribution in great detail, but a-priori assumed that the rise of the electron and phonon (lattice) temperature to be negligible (i.e., assumed they are both at room temperature)."

The phrase "i.e., assumed they are both at room temperature" mischaracterizes our paper [3]. The authors of Ref. [1] make a wrong comment about our paper [3], where it was assumed (as in many other papers working on the linear regime) that the lattice temperature is close to the room temperature, but the lattice temperature ***must not*** be equal to the room temperature, of course.

The phrase "[…] but a-priori assumed that the rise of the electron and phonon (lattice) temperature to be negligible" is also wrong. For calculations of [the] dynamic parameters, such as absorption and hot-electron rates, we can use the approximation $T_{lattice} \sim T_0$, but the physical picture requires $T_{lattice} > T_0$; again, the assumption for the linear regime is $T_{lattice} - T_0 \ll T_0$.

And, as also commented above, another mistake in Ref. [1], again, we did not introduce the electronic temperature in our paper [3], but used instead a more general model in the linear regime.

# 3. Cond-mat arXiv preprint (Ref. [1]), v3, page 10

In paper [1]: "The electron distributions we obtained are also different from the steady-state distributions obtained previously [21, 26-28]."

Here we highlight something that is not a critical comment about our work, but we would like to comment on it nonetheless. Ref. [1] does indeed present electron distributions that differ from our published results. But this is hardly surprising, given the fact that we have used fundamentally different models. Our study [3] is based on the fully quantized spectrum of electrons and incudes the dipolar and surface-assisted hot-electron generations, whereas Ref. [1] is based on the free electron approximation with no quantization.

# 4. Cond-mat arXiv preprint (Ref. [1]), v2, page 29.

A number of additional comments in this version of Ref. [1] include mistakes and misunderstandings of our work. Let's address the main ones independently:

**In [1]:** "[73] In fact, the numerical results in [27] *{Ref. [3] in this Commentary}* show that quantization effects are weak even for a 2nm particle!"

*Reply:* Quite the contrary. The quantization effects in our paper [3] are very strong and we stressed these effects. Below we will address this.

**In [1]:** "Indeed, the analytical result (red lines in figs 4 and 5 of [27] *{Ref. [3] in this Commentary}*) for the high-energy carrier generation rate, obtained by taking the continuum state limit, is very similar to the exact discrete calculation; peculiarly, this similarity in the results seem to be in contrast to their interpretation!"

*Reply:* In our paper [3], the interpretation and the results are in full agreement. The rates were obtained after the averaging over the nanocrystal size and, therefore, the size-quantization oscillations were smoothed (mimicking the case of a nanocrystal solution with size dispersion). Therefore, this comment in Ref. [1] seems to arise from the authors misunderstanding our work or not reading our paper carefully.

In our paper [3], the continuous approximation provides a reasonable estimate for the rate of generation of hot electrons (**averaged over sizes**), since we still have a large number of electrons in a nanocrystal (NC). The number of electrons in a NC $\gg 1$ and, therefore, it is reasonable to expect that the continuous approximation works well as an estimate for some quantities. Which is actually the case. Simultaneously, for the other spectral physical quantities, the quantization effects are very strong. In particular, strong quantization effects were found in the following physical properties: (1) the energy width of the low-energy generation rate in small NCs is dictated by the quantization; (2) the spectral hot-electron rates for a small NC with a given size oscillate with the energy strongly; to remove such oscillations, we performed the size averaging described above.

**In [1]:** "Moreover, the higher non-thermal electron generation in a dimer of spherical metals particles or in a cubic particle is seen to be associated almost completely with the higher average field enhancement and better resonance quality, both, purely classical aspects."

*Reply:* Again, this appears to be surprisingly disconnected from the actual contents of our paper [3], which focuses on the hot spot effects. In it, we state the presence of two enhancement mechanisms for the amplified quantum generation of hot electrons in confined plasmonic systems with hot spots: (1) the field-enhancement mechanism due to hot spots and (2) the quantum mechanism in hot spots of the cubes where the linear momentum of an electron is not conserved. Fig. 14 in Ref. [3] and Fig. 10 in Ref. [9] show that mechanism (1) is not able to explain the total increase in hot electron generation. Below we reproduce those figures. We note that our paper [9] has more data for the case of the dimer.

**Ref. [1] states** "[…] with the higher average field enhancement and better resonance quality, both, purely classical aspects".

*Reply:* The above aspects are indeed classical, but the surface-scattering mechanism of generation of nonthermalized hot electrons, which is the central topic of our paper [3], is an entirely quantum one, since the electrons in this mechanism absorb the photon quantum, $\hbar\omega$. Hot electrons with high energies are excited in [3] due to the surface scattering and the non-conservation of linear momentum at surfaces and in hot spots. The paper [1] is missing entirely the key surface-scattering and hot-spot mechanisms of hot electron generation, which come from the non-conservation of the linear momentum of an electron.

**In [1]:** "Note, however, that the interpretation of these results in [27] *{Ref. [3] in this Commentary}* was different. This shows that neglecting the possibility of momentum mismatch (which is the effective meaning of avoiding the energy state quantization, as essentially done in our calculations) provides a rather tight upper limit estimate."

*Reply:* The meaning of the comment above is unclear.

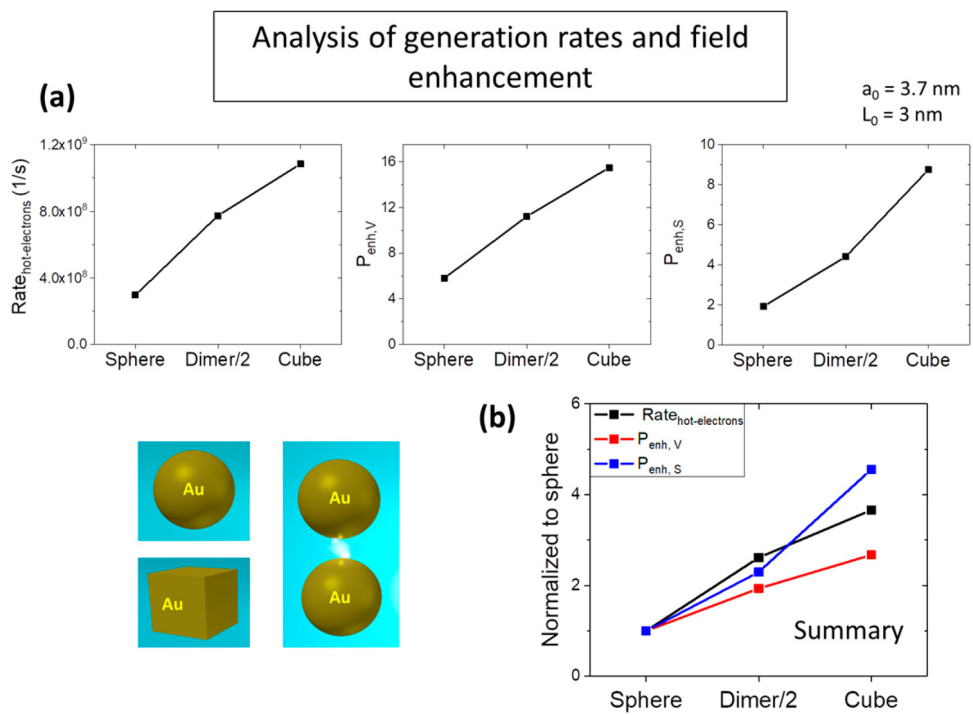

**Reproduction of Fig. 14 in Ref. [3].** Panel (b) illustrates how our theoretical results for the rates of generation of non-thermal hot electrons cannot be accounted for just by considering the field enhancement inside the NCs, or even the enhancement at their surfaces. Copyright American Chemical Society 2017.

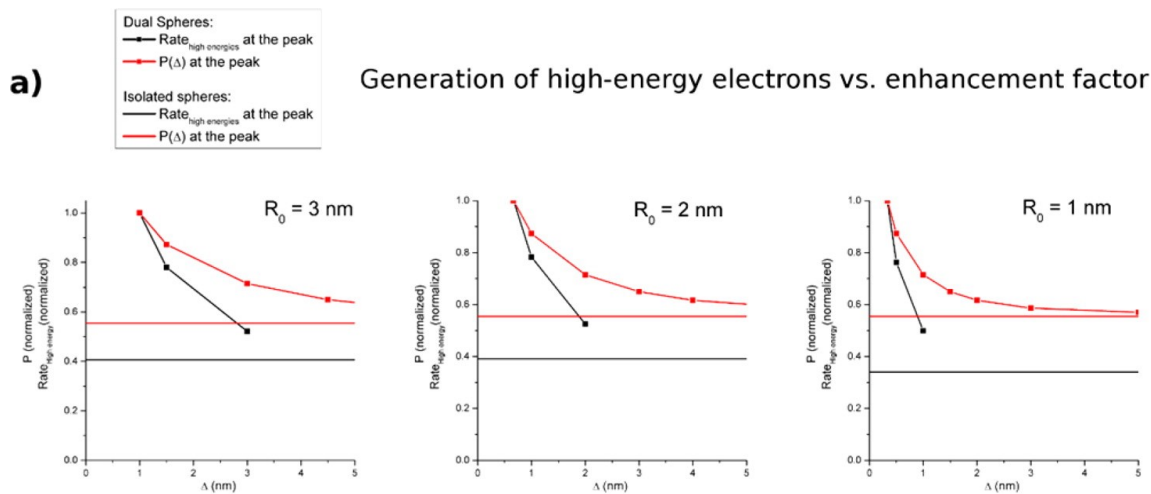

**Reproduction of Fig. 10 in Ref. [9].** Our theoretical results show that, by changing the gap size of the dimer, the growth of the rate of generation of non-thermal hot electrons is faster than the growth of the field enhancement in the volume of the NCs. Copyright American Chemical Society 2016.

# 5. Cond-mat arXiv preprint (Ref. [1]), v2, page 31.

**Paper [1] (version 2) reads:** "[84] Worse, in [27] *{Ref. [3] in this Commentary}*, the electron temperature was fixed arbitrarily to 1300K, whereas the single temperature (classical) calculation for this configuration shows that the temperature rise should be < 1K."

We again confront a strong mischaracterization of our work. This short, one-sentence comment contains 3 mistakes:

i) We did not introduce an electronic temperature ($T_e$) in Ref. [3], since that would represent a strong approximation. The electronic system is under CW driving and, strictly speaking, it is not described by a thermodynamic temperature. Although we should note that the electronic-temperature approach is in fact very convenient and it is widely used in the literature. In our paper [3], we simply did not use this approach of $T_e$, but an "effective temperature" ($T_{eff}$) for the distribution of electrons. Importantly, we never referred to it as an electronic temperature.

ii) $T_{eff}$ is a quantum parameter and it is not arbitrarily defined, but given by the computed width of the low-energy hot-electron distribution in a NC with a small radius, where the width of the electronic distribution is given by the quantum transition energy for the low-energy dipolar excitations. In our paper [3], $k_B \cdot T_{eff} = 0.1$ eV and it gives $T_{eff} = 1160$ K, value that was used for a small NC with a size of 4 nm (see Ref. [3], the text after Eq. (42)). The width of the electron distribution was calculated numerically and found to be 0.3 eV; this gives the effective temperature *via* the equation: Width = 0.3 eV = $3k_B \cdot T_{eff}$ (Fig. 4, right panel for $a = 4$ nm, in Ref. [3]). This width is a quantum value for a small NC and it can be estimated analytically as: Width ~ $v_F *(\hbar\pi/a)/2$ ~0.36 eV. The latter is, however, merely an estimate and it should not be considered to be precise.

iii) Lastly, again, we did not introduce the electronic temperature. The lattice temperature does depend on a variety of parameters that include the matrix and on other thermal details, and it is unclear from where did the authors of Ref. [1] obtain the value of 1 K, as we did not specify the relevant material parameters in our theory paper. In conclusion, we find this short comment strongly misleading and not at all relevant as a commentary on our paper.

Furthermore, in pages 21-22 (Ref. [1], v2), one can read: "In particular, we find that the efficiency of non-thermal electron generation is roughly independent of particle size, but the overall heating scales as $a^2$, in agreement with the single temperature (classical) heat equation. Such correspondence is absent in the simulations in [27] *{Ref. [3] in this Commentary}*, <u>where the temperature was not adjusted for particles of different sizes</u> [84] *{This is the footnote discussed in this section}.*"

The quantum physical mechanism for the hot-electron production used by us in Ref. [3] is absent in Ref. [1]. Then, a direct comparison between these works is not easy. The temperature increase in our paper [3] was taken as being small (the linear regime), and, therefore, we do not need to adjust this small quantity ($\Delta T$) for different NCs. The results presented in [1] for the excited-electron distributions and temperature are not novel conceptually in our opinion. The current literature has a very large number of papers with such effects.

# 6. Faraday Discussions (Ref. 2). Pages 7 (bottom) and 8 (top).

In Ref. [2], the authors argue that many of the theoretical accounts investigating the excitation of high-energy hot electrons and their relationship with photocatalytic effects are wrong in a fundamental sense. Again, here we will not task ourselves with the discussion of the scientific merit of the ideas presented in Ref. [2], but merely to address specific comments made in Ref. [2] about our papers, that we believe mischaracterize our published work.

Firstly, it is stated that our approach [3] does not describe excited electrons near the Fermi surface. This is clearly incorrect. Our approach does describe the non-equilibrium electron population under the CW illumination near the Fermi surface and, moreover, we stressed the important role of the low-energy electrons in the energetics of the plasmon. Regarding the thermal effects, which the authors state that they have been largely overlooked by the theoretical plasmonic community, we should note that they are integrated into our formalism, as it includes the lattice temperature of the NC as a parameter ($T_{lattice}$). In our calculations, this temperature is taken slightly above the ambient temperature and justified by working within the "linear-response theory" and "under weak illumination", as stated in Ref. [3]. Therefore, $T_{lattice} > T_0$ and the temperature increase is small, i.e. $\Delta T = T_{lattice} - T_0 \ll T_0$.

Furthermore, the authors of Ref. [2] also comment that our paper [3] "incorrectly accounts for the role of interband transitions". Given that this criticism is not developed, it is difficult for us to respond properly to it, but we think that we should make a couple of relevant remarks. Firstly, the goal of our work is the modelling of the **intraband** transitions that are responsible for the excitation of high-energy non-thermalized hot electrons, due to quantum surface scattering. It is our opinion that we made that clear in our manuscript [3]. Secondly, and with the previous comment notwithstanding, one can see that we did pay consideration to the **interband** transitions in Ref. [3] (e.g. see Figs. 4,5,S6) and accounted for them in terms of physically-justified models and approaches. Given these observations, the aforementioned comment in Ref. [2] seems gratuitous.

# 6. Light: Science & Applications, 8, 89 (2019) (Ref. 10).

**The paper [10] again contains several wrong and ill-defined statements about our work. We are under impression that the authors of [10] do not understand the content and terminology in our papers [3,4].** The paper [10] is similar to the preprint [1] but with some new additions. This section will comment on the additions in [10], whereas those are also in the preprint [1] were already addressed above.

## 6.1 Supporting Information in [10].

......................................................................

**Comment on p. 22, Supporting Information of [10]:** "[S55] In [S10] *{Ref. [3] in this commentary}*, the electron temperature was not evaluated self-consistently, as in our formulation, but rather, it was set by hand and referred to as an "effective" temperature; no discussion of the choice of values was given. Unfortunately, the effective electron temperature values were set to ~ 1300K (0.1 eV for a 4nm NP), whereas the single-temperature (classical) calculation for this configuration shows that the temperature rise should be ~ 0.13K. In addition, the scaling of the effective temperature used in [S10] violates the classical a^-2 scaling; in fact, it showed an inverse proportionality to the NP size (specifically, the effective temperature of a 24nm NP was ~ 520K (0,04eV)). Claims in [Govorov & Besteiro, ArXiv 2019] on the emergence of quantum effects in this context are questionable, due the relatively large size of the NPs studied in this case, see also the discussion at the end of Section I A."

**Again, the above comments are incorrect, since we did not introduce the electronic temperature at all in our paper [3]. It seems clear that the authors of [10] do not understand the meaning of the quantum parameter "effective temperature" in our paper [3]. Besides, the quantum effects for our NC sizes are strong since the quantization is stronger than the thermal energy.**
Again, we do not introduce "the electronic temperature" – we introduce "an effective temperature" for a small subset of excited electrons relaxed via the electron-electron relaxation mechanism. Those relaxed electrons are not thermalized. Our "an effective temperature" is fundamentally different to the term "an effective temperature" used in literature. Our parameter "an effective temperature" is a quantum parameter and governed by the quantization effect in a NC.

Our choice for the effective temperatures is fully justified and comes from our numerical results in [3]. Again, these quantum parameters were introduced for a small subset of electrons undergoing fast e-e relaxation:

$k_B T_{eff} = 0.1\,\text{eV}\ (1160\,\text{K})$ for $a_{NC} = 4$ nm
$k_B T_{eff} = 0.04\,\text{eV}\ (460\,\text{K})$ for $a_{NC} = 24$ nm

These energy parameters correspond exactly to the widths of the calculated spectra of excited low-energy electrons in Figure 7a,b (top panels). The widths of the low-energy peaks in Figure 7b,c [3] are:

$\text{Width} = 0.3\,\text{eV} = 3 \cdot k_B T_{eff}$ for $a_{NC} = 4$ nm
$\text{Width} = 0.12\,\text{eV} = 3 \cdot k_B T_{eff}$ for $a_{NC} = 24$ nm
(3c)

From the above widths, we obtain: $k_B T_{eff} = 0.1\,\text{eV}$ and $0.04\,\text{eV}$ for the chosen sizes of NCs. For our NC sizes, the quantum energy parameter $k_B T_{eff}$ should be higher than the thermodynamic energy due to ambient temperature ($k_B T_{room}$), due to the quantization of states in these small NCs (4 and 24 nm): $k_B T_{eff} \gg k_B T_{room}$, where $k_B T_{room} = 0.025\,\text{eV}$. We now estimate the quantization energy for a NC at the Fermi level. For simplicity, one can use a simple model of an electron in a box potential (plasmonic cube), having the following quantum spectrum:

$$E_{n_x,n_y,n_z} = \frac{\hbar^2 \pi^2 \left(n_x^2 + n_y^2 + n_z^2\right)}{2 m a_{NC}^2}.$$

For the electrons at the Fermi surface, the quantum numbers satisfy:

$$E_{n_x,n_y,n_z} = \frac{\hbar^2 \pi^2 \left(n_x^2 + n_y^2 + n_z^2\right)}{2 m a_{NC}^2} \approx E_F = \frac{m v_F^2}{2},$$

$$n_x^2 + n_y^2 + n_z^2 = \frac{E_F}{E_q},$$

$$E_q = \frac{\hbar^2 \pi^2}{2 m a_{NC}^2}.$$

Therefore, we observe that, at the Fermi surface, the quantum numbers satisfy:

$$\sqrt{n_x^2 + n_y^2 + n_z^2} = \sqrt{\frac{E_F}{E_q}}.$$

The electron quantum numbers, in occupied states, run independently between the limits:

$1 \leq n_x \leq n_F$

$1 \leq n_y \leq n_F$

$1 \leq n_z \leq n_F$

$n_F = \sqrt{\frac{E_F}{E_q}}.$

The above is valid for room temperature when $k_B T_{room} \ll E_F$. For an external electric field parallel to the x-axis, the largest dipolar transition near the Fermi level is $\Delta n_x = 1$. Therefore, we have for the optical transition energy:

$$\Delta E = E_{n_x+1, n_y, n_z} - E_{n_x, n_y, n_z} \approx \frac{\hbar^2 \pi^2 (2n_x)}{2 m a_{NC}^2},$$

$1 \leq n_x \leq n_F$

We assumed that $n_x \gg 1$. Then, we immediately see the range of optical excitation energies near the Fermi level:

$$0 < \Delta E < \Delta E_{max} = \frac{\hbar^2 \pi^2 n_F}{m a_{NC}^2} = \frac{\hbar \pi v_F}{a_{NC}}. \qquad (4c)$$

The width at half maximum (FWHM) of the distribution of excited low-energy electrons (like in Figure 7b,c in [3], extending through the Drude electrons with $E$ such that $E_F < E < E_F + \Delta E_{max}$) should be less than the total range of the interval, Eq. (4c) - so let's estimate it here as being half of the whole interval:

$$Width_{estimate} = \frac{\Delta E_{max}}{2} = \frac{1}{2} \frac{\hbar \pi}{a_{NC}} v_F.$$

From the above equation, we obtain:

$E_F = 5.6$ eV, $\quad v_F = 1.4 \cdot 10^8$ cm/s

$Width_{estimate} \sim 0.36$ eV for $a_{NC} = 4$ nm  (5c)

$Width_{estimate} \sim 0.06$ eV for $a_{NC} = 24$ nm

The above estimations are in good agreement with the numerical results for spherical NCs – see the data above in Eq. (3c). We note that, in Eq. (5c), the number for $a_{NC} = 24$ nm is smaller than the computed value (0.1 eV in Eq. (3c)) since the estimated width should be further corrected by including the thermal energy contribution $\sim k_B T_{room}$ (see Eq. (52) in [3]).

Of course, the above derivations provide us only with an estimate since we are using the approximation of a square box.

So, we again stress that the quantum effects for our NCs are strong since the quantization energy at the fermi level is much higher that the thermal energy:

$Width_{estimate} \gg k_B T_{room}$ for $a_{NC} = 4$ nm, 24 nm.

Finally, we should note that our paper [3] contains the discussion on the quantum origin of the width $Width_{estimate}$. In [3], the width of the distribution of the excited low-energy electrons is addressed in great detail: for small size, $Width_{estimate} \sim \Delta E_{crit}$, where the energy $\Delta E_{crit}$ is shown in Figure S3 and given by Eq. 52. We see that for small sizes: $\Delta E_{crit} \sim const \cdot \dfrac{\hbar \pi}{a_{NC}} v_F$. Therefore, the width of the excited low-energy electrons in our NCs with small sizes has clearly a quantum origin. **We are under impression that the authors of [10] did not read our paper with sufficient care.**

..................................................................................

**Comment on p. 16, Supporting Information of [10]**: "In particular, we find that the efficiency of non-thermal high-energy electron generation is independent of particle size, but the overall heating scales as a^2, in agreement with the single-temperature (classical) heat equation. Such correspondence is absent in the simulations in [S10] [S55]." **This comment is misleading and shows that the authors of [10] did not understand our paper [3].** Our paper investigates the hot-electron generation mechanism coming from the quantum surface-scattering channel, and our mechanism is not a classical mechanism of heating. The efficiency in our paper $\sim 1/a$ since we work with the surface effect.

........................................................................

**Comment on p. 17, Supporting Information of [10]:** "All these results indicate that unlike previous claims [S58] "quantum size effects" have at most a small quantitative effect on the non-thermal carrier generation efficiency. This result was corroborated in [S10], see discussion 17 at the end of Section I A."

**It is a wrong and misleading comment in [10]. Or paper [3] did not "corroborate" at all the above conclusions in [10]. This comments again indicates that the authors of [10] do not understand the context of our paper [3]. Our paper [3] tells clearly that the quantum effects are very important in small NCs, and they influence the hot-electron efficiency and the hot-electron distribution. The typical pattern of the authors of [10] – they are trying to use our results when they need, and simultaneously they are making many false comments on our work.**

........................................................................

**Supporting Information of [10]**: The text of supporting information of [10] is intricated and has multiple references to our papers and our approximations. To save time, we did not address the other references to our papers in [10], although those comments in [10] also have errors.

........................................................................

## 6.2 Main text of [10].

·······································································

**Long comment in the main text of [10], on p. 2:** "On the other hand, the few pioneering theoretical studies of the steady-state nonequilibrium under CW illumination [26, *{Ref. [4] in this commentary}* 27] account for the electron distribution in great detail, but assume that the electron and phonon (lattice) temperatures are both at room temperature. In [28] *{Ref. [3] in this commentary}* an "effective" electron temperature is referred to (without formally defining it); it is assumed to be higher than the environment temperature, but is pre-determined (rather than evaluated self-consistently). "

**This is a wrong comment, and it was already address above.** Page 2762 in our paper [3] introduces explicitly the lattice temperature: "As one can see from Figure 2a, a plasmon decays into high- and low-energy electrons and holes in the Fermi sea. In the following step, these hot and warm electrons emit phonons and locally increase the lattice temperature".

The "effective temperature", defined for a small subset of the electrons subject to relaxation through e-e scattering, is a quantum parameter in our study, and this parameter is fundamentally different to the thermodynamic *T*. The effective *T* in small NCs is given by the size-quantization energy. The effective *T* in our study [3] was discussed by us above in great detail.

**Comment on p. 2 in [10]:** "As discussed in SI Section S3, the chosen values for that "effective" electron temperature are questionable and while it is claimed in [29] that Tph > Tenv, there is no indication of this in the published papers."

**This comment is misleading, as addressed above. We did not use the parameter "the electronic temperature" in our study, and this comment is made under the assumption that we did. Besides, the statement $T_{ph} > T_{env}$ is clearly made in our original paper [3] on p. 2762.** Simultaneously, as we investigate the excitation of hot carriers in a linear optical regime where $(T_{ph} - T_{env})/T_{env} \ll 1$, so that $T_{ph} \approx T_{env}$.

·······································································

**Another comment on p. 2 in [10]:** "All these results are very different from those known for electron dynamics under ultrafast illumination, as well as from previous studies of the steady-state scenario that did not account for all three energy channels (e.g.,26–28,40)."

**This comment is false again.** Our study included all main channels of excitation and dissipation: Optical driving, e-e relaxation, and e-phonon relaxation. In contrast to [10], our study is based on the quantum theory, and our results are fully consistent to the quasi-classical results (Boltzmann equation) in the appropriate limiting cases.

• • • • • • • • • • • • • • • • • • • • • • • • • • • • • • • • • • • • • • • • • • • • • • • • • • • • • • • • • • • • • • • • • • • • • • • • • • • • • • • • • • • •

Page 5: "As a result, quantitative conclusions on the distribution drawn in these studies are incorrect. Similar inaccuracies are also found in the calculations of [21,26–28]."

**Again, it is a wrong comment in [10]. All our quantitative conclusions in [3] are correct within our approach and are fully applicable to experiments, and, in fact, our results [3] have been successfully applied in several recent experiments, guiding experimental work in the field of research on hot electrons.**